\documentclass[aip,twocolumn,reprint]{revtex4-1}

\def\rtx@aipapl{%
\typeout{Using journal substyle \@journal.}%
\@booleanfalse\authoryear@sw
\let\@runningtitle\@empty
}%

\usepackage{amsmath}
\usepackage{graphicx}
\usepackage{color}

\begin{document}
		

\title{Wideband Excitation of Microbubbles with Chirps to Maximize the Sonoporation Efficiency and Contrast in Ultrasound Imaging}
\author{Sevan~Harput}
\author{James~McLaughlan}
\author{David~M.~J.~Cowell}
\author{Steven~Freear}
\affiliation{Ultrasonics and Instrumentation Group, School of Electronic and Electrical Engineering, University of Leeds, Leeds, LS2~9JT, UK.}


\begin{abstract}

The importance of the excitation bandwidth is well known in diagnostic ultrasound imaging. However, the effect of excitation bandwidth in therapeutic applications of microbubbles has been mostly overlooked. A majority of contrast agent production techniques generate polydisperse microbubble populations, so a wide range of resonance frequencies exist. Therefore, wideband excitation is necessary to fully utilize microbubble resonance behavior and maximize the reradiated energy from a microbubble population, both for imaging and therapy. 

Oscillations of sixty SonoVue microbubbles in proximity of a rigid boundary were captured on a high speed camera at 3 Mfps, excited with a peak negative pressure of 50 kPa at 1 MHz. Measurements were analyzed according to their peak radiated pressure, radial oscillations, root mean squared pressure, and shear stress generated by microbubbles. Results showed that long duration and wideband excitation at low intensity levels was preferable for sonoporation, where microbubbles can be driven in a stable oscillation state without experiencing inertial cavitation or destruction.

\end{abstract}


\maketitle

\section{Introduction}
\vspace*{-2mm}
Microbubbles (MBs) are used predominantly as ultrasound contrast agents for imaging applications; \textit{e.g.} echocardiography~\cite{Yeh2015}, perfusion and Doppler imaging~\cite{Tremblay-Darveau2014}, and molecular imaging of cardiovascular diseases~\cite{Lindner2009}. The therapeutic potential of MBs is mostly seen as a delivery mechanism such as sonoporation of endothelial cells to facilitate drug delivery~\cite{Kooiman2011,Skachkov2014}, enhancing DNA transfection using sonoporation~\cite{Greenleaf1998,Qiu2012,Delalande2013}, and localised drug delivery using acoustic radiation force~\cite{Raiton2012}.

Sonoporation of the cell membrane is generally pronounced as one of the main mechanisms that improves the cellular uptake~\cite{Bao1997,Delalande2013}. Release of a drug can be spatiotemporally controlled by using ultrasonically triggered encapsulated MBs, which interact strongly with an ultrasonic field and oscillate. There are several hypothesis regarding to sonoporation and increasing the efficacy of ultrasound induced drug delivery. According to Escoffre \textit{et al.} and Meng \textit{et al.} MB destruction is a key parameters in sonoporation efficiency~\cite{Escoffre2013,Meng2014}. Karshafian \textit{et al.} suggests that MB disruption is a necessary but an insufficient indicator of sonoporation~\cite{Karshafian2009}. However, it has been presented that cell viability decreases with the increasing acoustic pressure~\cite{Karshafian2009,Qiu2012}. The use of high power ultrasound, which leads to instantaneous bubble destruction, is causing concerns about biosafety of this technique~\cite{Yeh2015}. Stable cavitation, mechanical streaming and radiation forces are the main non-thermal effects obtained at low acoustic pressure levels, which makes the low-intensity reparable sonoporation a suitable candidate for therapeutic delivery~\cite{Delalande2013}. Therefore, a great deal of recent studies focus on maximizing the cell viability by using low intensity ultrasound~\cite{Nejad2011,Qiu2012,Skachkov2014}.

A MB should be excited at its natural frequency to achieve sonoporation at low intensities, where they absorb and therefore reradiate more acoustic energy. This localized energy may be used to target specific cell populations in the vasculature with minimal impact on the surrounding tissues. Resonant MBs also improve the separation between tissue and contrast agents with the application of contrast-specific methods such as pulse inversion and amplitude modulation and therefore resulting in a higher contrast-to-tissue ratio~\cite{Harput2012,Harput2013,Harput2013a,Harput2013b}. Most commercial phospholipid encapsulated MBs typically have a polydisperse size distribution (1-10~$\mu$m), where a range of resonant frequencies would exist in a single population. The response of a polydisperse MB population to wideband and long duration excitation is stronger~\cite{Harput2013a,McLaughlan2013}, since more MBs can be excited close to their resonance frequency.

\section{Materials and Methods}
\vspace*{-2mm}
To evaluate the effect of bandwidth, a high speed imaging setup was created to record the oscillations of MBs next to a boundary. The rigid boundary was decided to be the most relevant choice for the drug and gene delivery with targeted MBs and molecular imaging, because the development of a tumor can increase the stiffness of the vasculature and interstitial tissue, which will result in a rigid vessel wall~\cite{Qin2007}. 

 \begin{figure}[!t]
 	\centering
 	\includegraphics[viewport = 0 0 401 457, width = 70mm, clip]{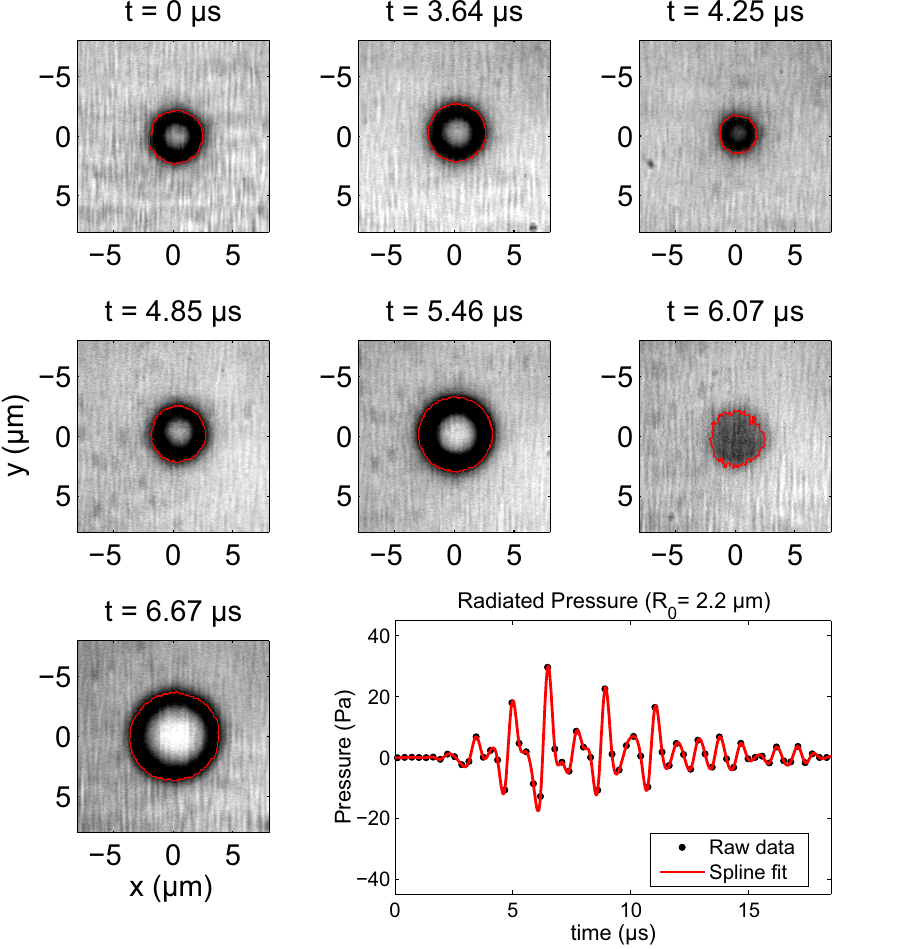}
 	\caption{First 7 sub-figures show high speed camera frames captured at 3.3 Mfps. The red contour represents the perimeter of the MB. The radiated pressure from a MB with 2.2~$\mu$m radius calculated by Eq.~\eqref{eq:RPpres} is shown in the last sub-figure.  }
 	\label{fig:Oscillation_frames_combined}
 \end{figure}

Three different waveforms were used during the experiments and the results were analyzed by using 4 different metrics; root mean squared (RMS) pressure, peak radiated pressure, radial oscillations, and shear stress generated by MBs. A linear frequency modulated (LFM) chirp was generated with a center frequency of $1$~MHz, duration of 15~$\mu$s, and fractional bandwidth of 75\%. For comparison, a sinusoidal tone-burst was generated with a center frequency of $1$~MHz, and a duration of 15~$\mu$s, and a Gaussian pulse was generated with a center frequency of $1$~MHz, and a fractional bandwidth of 75\%. The reasons for choosing these waveforms were that the direct comparison of the LFM and the tone-burst reveals the effect of bandwidth, and the direct comparison of the LFM and the Gaussian pulse reveals the effect of signal duration on MB oscillations. Among these excitation signals LFM with the higher time-bandwidth product will give the highest SNR for imaging~\cite{Misaridis2005,Harput2011}.

A $1$~MHz V303-SM immersion transducer (Olympus-NDT Inc., Waltham, MA) with 75\% fractional bandwidth was calibrated to generate an \textit{in situ} peak negative pressure of $50$~kPa in a $\mu$-Slide I $0.4$ (ibidi GmbH, Munich, Germany), where MBs from SonoVue (Bracco S.p.A, Milan, Italy) were individually imaged. The $\mu$-Slide was placed on a Nikon Eclipse Ti-U inverted microscope (Nikon Corp., Tokyo, Japan), which was coupled to Cordin 550 high speed camera (Cordin Company, Salt Lake City, Utah) working at $3 \pm 0.3$ Mfps.

Oscillations of 60 MBs next to a rigid boundary were imaged with the described experimental setup above. A total of 4 measurements were discarded because of image noise, MB aggregation, and MB destruction (2 measurements). The remaining 56 recorded high speed measurements consisted of 10 Gaussian pulse, 22 sinusoidal tone-burst, and 24 LFM chirp excitations. Peak and bandwidth of the signals were measured by using linear interpolation ~\cite{Harput2014a}.

To estimate the acoustic pressure radiated from individual MBs, radial oscillations of MBs were calculated from the high-speed camera frames as shown in Fig.~\ref{fig:Oscillation_frames_combined}. First, the area of MBs were measured by using a global image thresholding method~\cite{Otsu1979} in Matlab (The MathWorks Inc., Natick, MA). The area of the MB, as highlighted with a red contour in Fig.~\ref{fig:Oscillation_frames_combined}, was later used to calculate the bubble radius, $R$, with an assumption of a circular 2D MB footprint; $R = \sqrt{\text{Area} / \pi}$. Finally, the emitted pressure, $P(t)$, in a liquid with the density of $\rho$ was calculated at a distance of $d$ from the bubble center as~\cite{Morgan2000}
\begin{equation}
P(t) = \frac{\rho}{d} ( {R^2}\ddot{R} + 2R \dot{R}^{2} ).
\label{eq:RPpres}
\end{equation}

\section{Results and Discussion}
\begin{figure}[!t]
  	\centering
  	\includegraphics[viewport = 30 0 485 300, width = 88mm, clip]{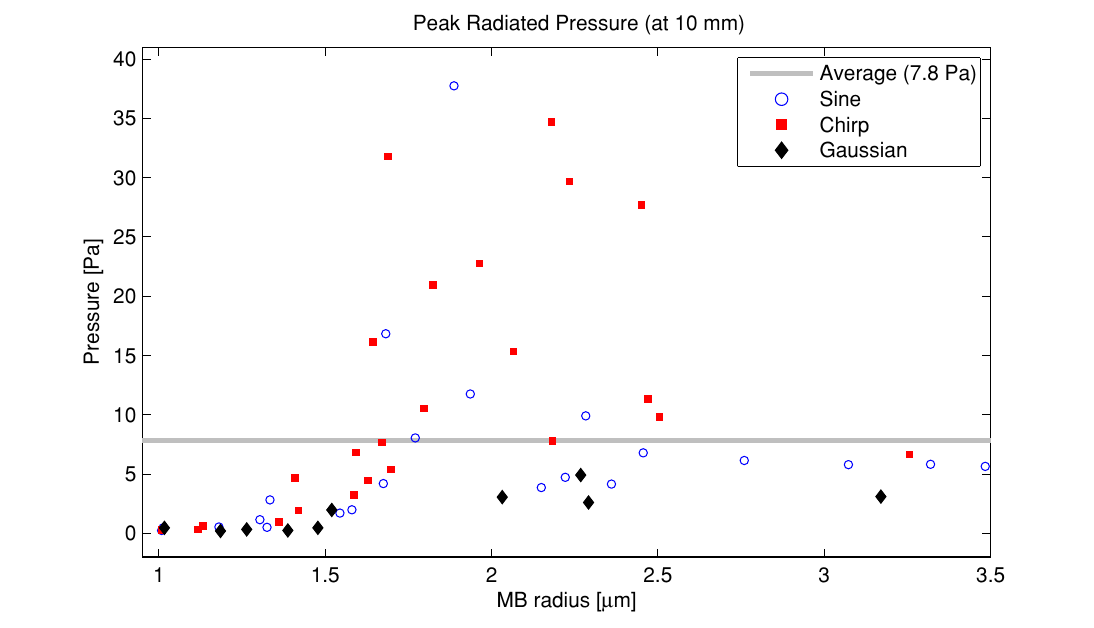}
  	\caption{The peak pressure radiated from MBs excited by sinusoidal tone-burst (blue circles), chirp (red squares), and Gaussian pulse (black diamonds) excitations. The gray line is the average of all data points for all excitations, which is $7.8$~Pa. }
  	\label{fig:peak_rad_pressure_with_avg}
\end{figure}

\vspace*{-2mm}
Peak pressure is one of the most significant metrics for contrast-enhanced ultrasound imaging with targeted or un-targeted MBs, since image contrast is directly proportional with the pressure. Fig.~\ref{fig:peak_rad_pressure_with_avg} shows the peak pressure values calculated at $d=10$~mm and their average value. The peak radiated pressure was defined as the largest of peak negative pressure or peak positive pressure; $\textbf{max}(|P^-|,|P^+|)$.

None of the MBs excited by a Gaussian pulse has generated a peak pressure value above the average of $7.8$~Pa due to the low energy content of the pulsed excitation. Long duration waveforms have generated higher peak pressure values, where 23\% (5 out of 22) of sinusoidal tone-burst excitations, and 46\% (11 out of 24) of LFM chirp excitations were above the average pressure level. Although long duration excitation has the potential to increase the image contrast, it results in poor image resolution. For sinusoidal tone-burst excitation, the resolution cannot be improved; however, a matched filter will recover the resolution of a chirp signal by compressing the scattered energy with a certain chirp rate into a single pulse~\cite{Harput2014}. Therefore, chirp excitation is the most suitable excitation method for contrast-enhanced ultrasound imaging~\cite{Harput2015}.

Although, the emitted acoustic pressure is an effective method to analyze the MB behavior, the peak radiated pressure is not an ideal metric to evaluate the MB response for both imaging and therapy. Sonoporation can be defined as the transient increase in the cell membrane permeability and porosity; however it is difficult to define a sonoporation threshold based on the ultrasonic exposure parameters. Greenleaf \textit{et al.} observed the gene transfection threshold at a spatial average peak positive pressure of 0.12 MPa at 1 MHz with Albunex MBs~\cite{Greenleaf1998}. Wu \textit{et al.} shown that the shear stress threshold for sonoporation generated by microstreaming was determined to be $12 \pm 4$~Pa at $21.4$ kHz~\cite{Wu2002}. Kooiman \textit{et al.} reported that targeted lipid-coated MBs induced drug uptake in endothelial cells when their relative vibration amplitude was larger than $50\%$ for a peak negative pressure as low as 80~kPa at 1~MHz~\cite{Kooiman2011}. Therefore, the shear stress values and the radial oscillations of MBs were calculated to assess their sonoporation efficiency.

The shear stress generated by an oscillating MB can be defined as~\cite{Wu2002}
\begin{equation}
S = \frac{ \sqrt{\rho \omega^3 \eta } \> (R-R_0)^2 }{\sqrt{2} \> R_0} ,
\end{equation}
where $\rho$ is the density of liquid, $\eta$ is the viscosity of liquid, $R_0$ is the initial bubble radius, and $\omega$ is the angular frequency.

After analyzing all 56 measurements, a maximum shear stress value of 31~kPa and a minimum shear stress value of 35~Pa were calculated with an average of 6~kPa. The average shear stress generated by Gaussian pulse, sinusoidal tone-burst, and chirp excitations were $1.1$~kPa, $3.3$~kPa, and $10.4$~kPa respectively. These values indicate that it is possible to achieve sonoporation for all excitation methods at as low as 50~kPa peak negative pressure, where MB destruction is observed in only 3.3\% of all measurements. Chirp excitation is generated 3 and 9 fold higher stress compared to sinusoidal tone-burst and Gaussian pulse excitations on average, and therefore it is the most effective excitation method used in this study.

\begin{figure}[!t]
  	\centering
  	\includegraphics[viewport = 30 0 485 300, width = 88mm, clip]{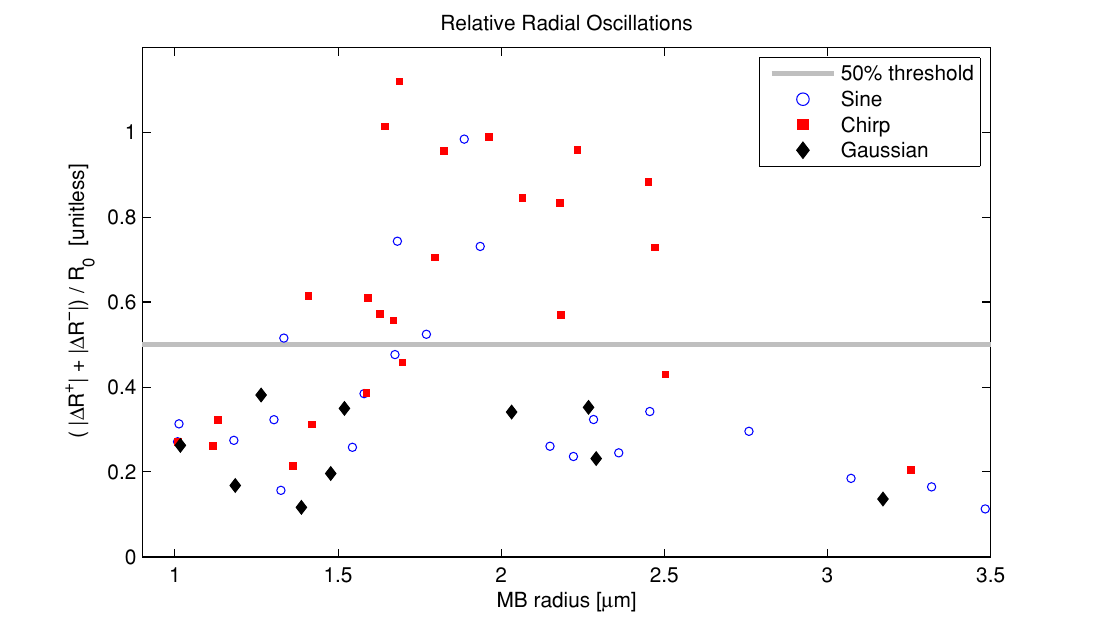}
  	\caption{The relative radial oscillations generated by MBs excited by sinusoidal tone-burst (blue circles), chirp (red squares), and Gaussian pulse (black diamonds) excitations. The gray line is the 50\% threshold, which is determined by~\cite{Kooiman2011}. }
  	\label{fig:rel_rad_oscillations}
\end{figure}  
  
Fig.~\ref{fig:rel_rad_oscillations} shows the relative radial oscillations, which is the sum of the maximum radial expansion and compression; $(|\Delta R^+| + |\Delta R^{-}|) / R_0 $. Analysis of relative radial oscillations resulted in similar conclusions with the peak pressure analysis. None of the MBs excited by a Gaussian pulse has generated radial oscillations above the 50\% threshold. Most of the MBs (63\%) excited by a chirp have generated larger radial oscillations than 50\%. For sinusoidal tone-burst excitation, however, only 23\% of MBs have generated more than 50\% radial oscillations.

The peak radiated pressure, shear stress, and relative radial oscillations are metrics that based on peak values, where the temporal effects are disregarded. Therefore, the root mean squared (RMS) pressure is used to summarize the general MB behavior as

\begin{equation}
P_{\text{rms}} = \sqrt{ \frac{1}{T_2-T_1} \int\limits_{T_1}^{T_2} [P(t)]^2 dt} ,
\label{eq:RMSpressure}
\end{equation}
where $T_1$ and $T_2$ are the time interval, and $P(t)$ is the pressure calculated by Eq.~\eqref{eq:RPpres} at $d=10$~mm.

\begin{figure}[!t]
  	\centering
  	\includegraphics[viewport = 30 0 485 300, width = 88mm, clip]{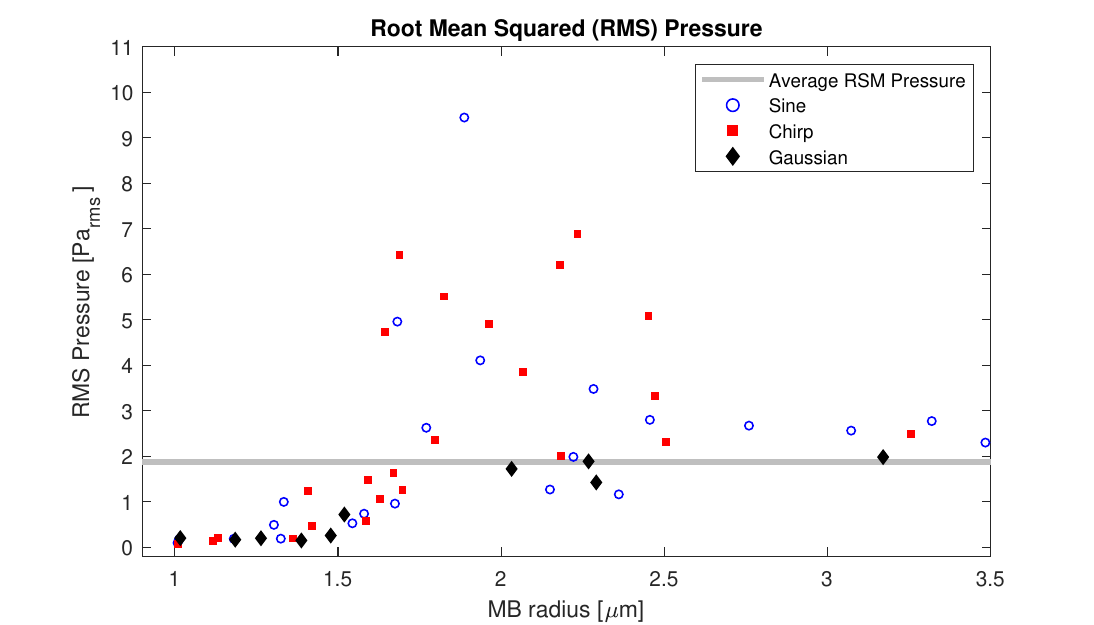}
  	\caption{The root-mean-squared (RMS) pressure values calculated by Eq.~\eqref{eq:RMSpressure} for sinusoidal tone-burst (blue circles), chirp (red squares), and Gaussian pulse (black diamonds) excitations. The gray line is the average of all data points for all excitations, which is $1.87$~Pa$_{rms}$. }
  	\label{fig:RMSpressure}
  \end{figure}

Fig.~\ref{fig:RMSpressure} shows the scattered RMS pressure from MBs. The resonant MBs were between $1.55$~$\mu$m and $2.7$~$\mu$m radius for chirp and Gaussian pulse excitations, and $1.675$~$\mu$m and $2.0$~$\mu$m radius for the sinusoidal tone-burst excitation. 

The tone-burst excitation has the largest peak amongst all excitations in Fig.~\ref{fig:RMSpressure}, but the narrowband sinusoidal waveform will efficiently drive only a small subpopulation of polydisperse MBs. Even for monodisperse MBs, the resonance frequency shifts as a function of distance between the MB and arterial wall~\cite{Qin2007}. Therefore, it is difficult to maximize therapeutic efficiency with MBs using a narrowband excitation, which is also not suitable for imaging. The response from the MB population when excited by a Gaussian pulse, shown in Fig.~\ref{fig:RMSpressure}, is weaker than the chirp excitation due to its lower energy content. The chirp excitation generates the ideal microbubble response for imaging compared to the other methods thanks to high pressure generation for a wide range of MBs. There is a smaller deviation between the peak and average RMS pressure values for chirp excitation that makes it possible to control the behavior of a larger portion of the MB population.


\section{Conclusions}
\vspace*{-2mm}
The resonance behavior increases the radiated pressure from MBs at a specific frequency that can be beneficial in both diagnostic and therapeutic ultrasound. However, dynamics of MBs change when they are injected into the blood stream. For a rigid boundary, the natural frequency shifts toward the lower frequency region when the distance between the bubble and the boundary is reduced. Near to an elastic boundary, the MB natural frequency can both decrease and increase~\cite{Doinikov2011}. Therefore, to trigger the resonance behavior for a larger amount of the MB population, a wideband excitation is necessary.

The resonance behavior precipitates strong MB oscillations, which can result in inertial cavitation and cause cell death at high pressure levels. Therefore, therapeutic ratio, which defined as the ratio of permeabilised to nonviable cells, is used to measure the sonoporation efficiency~\cite{Karshafian2009}. Although there are opposing views regarding to sonoporation, therapeutic efficiency is usually inversely proportional with cell death, which can be prevented by using low intensity ultrasound. The acoustic energy lost by reducing the intensity can be compensated by increasing the duration of the excitation. Findings of Nejad \textit{et al.} showed that inertial cavitation, which causes lethal sonoporation, is observed in a microsecond time scale and low intensity ultrasound sonoporation happened on a time scale of millisecond~\cite{Nejad2011}, which also favors the use of long duration waveforms. Therefore, this study concludes that wideband and long duration excitation at low intensity levels is preferable for sonoporation, where MBs can be driven in a stable oscillation state without experiencing inertial cavitation.

\section*{Acknowledgment}
\vspace*{-2mm}     
This work was supported by EPSRC grant EP/K029835/1. J. M. would like to acknowledge Leverhulme fellowship ECF-2013-247. The Cordin 550 high speed camera was borrowed from EPSRC Engineering Instrument Pool (Loan 3793).


\bibliography{references}       
\bibliographystyle{aipnum4-1}

\end{document}